# Investigation of Magnetic Proximity Effect inTa/YIG Bilayer Hall Bar Structure

Yumeng Yang[1,2], Baolei Wu[1], Kui Yao[2],Santiranjan Shannigrahi[2], Baoyu Zong[3], Yihong Wu[1, a)]

[1]*Department of Electrical and Computer Engineering, National University of Singapore, 4 Engineering Drive 3, Singapore 117583*

[2]*Institute of Materials Research and Engineering, A*STAR (Agency for Science, Technology and Research), 3 Research Link, Singapore 117602*

[3]*Temasek Laboratories, National University of Singapore, T-Lab Building, 5A, Engineering Drive 1, #09-02, Singapore 117411*

In this work, the investigation of magnetic proximity effect was extended to Ta which has been reported to have a negative spin Hall angle. Magnetoresistance and Hall measurements for in-plane and out-of-plane applied magnetic field sweeps were carried out at room temperature. The size of the MR ratio observed (~$10^{-5}$) and its magnetization direction dependence are similar to that reported in Pt/YIG, both of which can be explained by the spin Hall magnetoresistance theory. Additionally, a flip of magnetoresistance polarity is observed at 4 K in the temperature dependent measurements, which can be explained by the magnetic proximity effect induced anisotropic magnetoresistance at low temperature. Our findings suggest that both magnetic proximity effect and spin Hall magnetoresistance have contribution to the recently observed unconventional magnetoresistance effect.

[a)] Author to whom correspondence should be addressed: elewuyh@nus.edu.sg

## I. INTRODUCTION

Platinum (Pt) has been investigated as a promising spin current detector in recent reports on spin pumping effect[1-5] and spin Seebeck effect[6-8] involving ferromagnet (FM)/non-ferromagnet (NM) structures. In these investigations, the generated spin current is converted into an electromotive force via inverse spin Hall effect (ISHE), which can then be detected electrically. On the other hand, recently anisotropic-magnetoresistance (AMR)-like effect in Pt on the insulating ferromagnet yttrium iron garnet (YIG) was reported by Weiler *et al.*[9] and Huang *et al.*[10], which is attributed to the induced magnetic dipole moment in Pt due to magnetic proximity effect (MPE). This "pseudo-magnetization" in Pt was later supported by X-ray magnetic circular dichroism (XMCD) measurement[11]. Further investigation by Nakayama *et al.*[12] found that the magnetization direction dependence of this resistance change is distinct from any of the known MR effect. They proposed a spin Hall magnetoresistance (SMR) theory[13] to explain the experimental observations by taking into account the influence of spin current to the charge resistance through spin-orbit coupling (SOC). Some of the follow-up works[14-19] are in supportive of this model, in the way that the governing parameters such as spin mixing conductance of the interface, spin diffusion length and spin Hall angle of Pt could be extracted and appeared to be comparable to those of previous studies.

Although the presence of unconventional MR effect is confirmed, its underlying mechanism remains to be unveiled. Contradictory XMCD results showing the presence of only negligible Pt magnetic polarization in Pt/YIG structure[20,21] cast some doubts on the strength of MPE. On the other hand, the incompleteness of SMR theory was also indicated by the absence of similar effect in Au/YIG[22] and its preservation in Pt/surface modified YIG[23] structures. Therefore, in order to reveal the true mechanism of spintronic effects at FM/NM interfaces, it is necessary to extend the study to different FM/NM material systems. So far, with most investigations focusing on Pt/YIG, only a few have extended the bilayer structure to Au/YIG[22], Ta/YIG[17], Pt/Py[24], Py/YIG[23], $Fe_3O_4$/Pt, $NiFe_2O_4$/Pt[14], and $CoFe_2O_4$/Pt[19].

In this work, we focus on the Ta/YIG interface and discuss for the first time the Hall measurement results. The MR effect observed at room temperature is similar to that observed in Pt/YIG. The out-of-plane Hall resistance was found to be influenced by YIG substrate's magnetization at magnetic field below its saturation value of around 2000 Oe. Both observations could be explained by the SMR theory

However, we observed an additional dip centered at zero-field between around -200 Oe and 200 Oe, which could not be attributed to YIG's in-plane magnetization rotation[15]. The shape of the in-plane Hall resistance is different from that predicted by the SMR theory nor the MPE-induced planar Hall effect (PHE), but instead similar to that of the MR curve. The additional dip and shape difference might be due to the possible influence from MR. In the MR measurement with out-of-plane magnetic field, a flip of the MR polarity was observed at 4 K, implying the enhancement of MPE at low temperature. The results of the temperature dependent Hall measurements were similar to those of the room temperature measurements. Our findings suggest that both the SMR and the MPE have a contribution in the recently observed MR effect.

## II. SAMPLE FABRICATION AND CHARACTERIZATION

Samples were fabricated on (111) oriented YIG (8 μm)/ GGG (500 μm) single crystal wafer with a size of 5mm × 5mm. Fig. 1(a) shows the X-ray diffraction (XRD) pattern of an unpatterned reference Ta thin film (10 nm) deposited on YIG/GGG substrate. Comparison with the XRD pattern of bare substrate (inset of Fig. 1(a)) confirms that the Ta layer is in β-phase. To fabricate the Hall bars, the substrates were properly cleaned with acetone and isopropyl alcohol before coated with a Microposit S1805/PMGI SF6 bilayer resist. The resist-coated substrate was subsequently exposed using a Microtech laser writer to form a Hall bar pattern. The size of the central area of the Hall bar is 0.2 mm × 2.3 mm, and that of the transverse electrodes is 0.1 mm × 1 mm. A Ta (5 nm, 7 nm, 10 nm) layer was subsequently deposited using a sputter with a base pressure below $5\times10^{-5}$ Pa, and in an Ar working pressure of $5\times10^{-1}$ Pa, followed by liftoff to form the Hall bar. Before electrical measurements, the as-fabricated samples were bonded to chip carriers using a wire bonder. The resistivity of the samples (5 nm, 7 nm, 10 nm) falls into the range of $6.6\times10^{2}$ to $1.1\times10^{3}$ μΩ cm, which is in good agreement with the reported values for β-phase Ta[17].

Figs. 2(a)-(b) and Figs. 3(a)-(b) are the schematic configurations for MR and Hall measurements. For room temperature electrical measurements, the samples were placed in the ambient environment with in-plane (*x*, *y* directions) and out-of-plane (*z* direction) magnetic field applied, respectively. Low temperature measurements from 4 K to 250 K were performed in a variable temperature cryostat. A standard lock-in technique with 100 μA, 163 Hz AC current was used for the MR measurement, while the Hall measurement was performed using a DC technique with a current of 100 µA. The hysteresis loops for YIG/GGG were measured using vibrating sample magnetometer (VSM) as a reference to assist the interpretation of electrical transport data.

## III. RESULTS AND DISCUSSION

### A. Magnetic properties of YIG

Fig. 1(b) shows the in-plane and out-of-plane M-H loops for YIG/GGG substrates. The observation of an in-plane saturation field below 100 Oe and out-of- plane saturation field of around 2000 Oe suggests an in-plane anisotropy for (111)-oriented YIG/GGG. The inset shows a superimposition of the in-plane M-H loop and the M-H loop obtained with a 90° in-plane rotation of the sample with respect to its original position. The close overlap of two loops suggests no preferred easy axis in the (111) plane. Notably, for both cases, the coercivity ($H_c$) for YIG is below 10 Oe, which is reasonable for a ferrimagnetic material.

### B. Room temperature MR and Hall measurements

Figs. 2(c), (e) and (g) are the MR (hereafter referred to as the MR observed in the measurements of this work regardless of its origin) ratio of Ta (5 nm)/YIG for three applied magnetic field directions ($H_x$, $H_y$, $H_z$), respectively. Conventionally, for NM, its resistance can increase slightly as the applied field increases due to the Lorentz force felt by the conduction electrons[25]. This effect is named as ordinary magnetoresistance (OMR) with a positive polarity. In the present case, since the field applied is small (maximum 500 Oe), OMR is negligible in the field sweeps. However, a dip or peak is observed during all sweeps in the field region between -50 Oe and +50 Oe, indicating the presence of interfacial coupling in Ta/YIG. The MR ratio of $2\times10^{-5}$ is comparable to that reported in Pt/YIG[10]. If we only focus on the MR in *x* and *y* directions, it seems that it can be explained by the proximity effect, i.e., the anisotropic magnetoresistance (AMR) arises from the magnetized Ta layer (hereafter named as MPE-induced AMR). However, if this is the case, the MR in *z* direction should be negative as well instead of being positive as observed experimentally. Instead of the MPE-induced AMR, the experimental data can be understood as follows using the SMR theory.  According to this theory, the MR is determined by the angle between spin polarization of electrons in the Ta and YIG's magnetization direction. In the present case, as current is applied in *x* direction, the spin polarization induced by SOC should be dominantly in y direction. In this sense, the resistance response to applied field in *x* and *z* direction should be similar except that the MR

curve in *z* direction is broader than that in *x* direction due to shape anisotropy of the YIG layer.

Figs. 2(d), (f) are the Hall resistance of Ta (5 nm)/YIG for in-plane applied magnetic field directions ($H_x$, $H_y$), respectively. Conventionally, for in-plane field, the Hall effect in NM should vanish with only the presence of a background offset resistance coming from electrodes asymmetry. Similar to the longitudinal resistance, we observed a dip or peak in the in-plane Hall resistance located at around -50 Oe and +50. The magnitude of around 10-20 mΩ is on the same order of those reported by N. Vlietstra *et al.*[15]. However, its shape is different from that predicted by SMR or MPE-induced PHE, which should be an odd function of the applied magnetic field. One possible reason is that due to the relatively large size of the pattern and current distribution in the electrodes, the contribution from the longitudinal MR to the Hall resistance cannot be excluded. As the longitudinal MR is one order of magnitude larger than the Hall resistance, it is possible that the MR signal covers the true Hall signal. This can also be inferred from the similarity between the shape of the MR and Hall curves.

The out-of-plane Hall resistance of Ta (5 nm)/YIG, as shown in Fig. 2(h), is different from the linear ordinary Hall effect for NM. It has a nonlinear region between -2000 Oe and 2000 Oe, which coincides with YIG's saturation field. This is presumably caused by the fact that the total field experienced by electrons inside Ta is the sum of externally applied field and the stray field from the YIG. The latter is large when the in-plane magnetization of YIG is oriented into the vertical direction by an external field, while it is relatively small when it is saturated in-plane. The additional stray field dominates in the field range below saturation, causing the nonlinear region. While, when the field is above the saturation, the contribution from the applied field dominates, resulting in the linear region. Noticeably, an additional dip is observed in the center between -200 Oe and 200 Oe. N. Vlietstra *et al.*[15] related it to the in-plane magnetization rotation at $H_c$, which is unlikely the origin of the dip observed here as the $H_c$ of YIG is below 10 Oe, as shown in the M-H loop. Considering its similar shape and field range as that of the out-of-plane MR curve in Fig. 2(g), we believe that it is also related to the influence of MR signal as discussed above.

### C. Temperature-dependent out-of-plane MR and Hall measurements

So far, it seems that most of our results at RT can be explained by the SMR theory. However, we still cannot completely exclude the MPE contribution in Ta/YIG bilayers, as the proximity effect may be enhanced at low temperature. To this end, temperature-dependent MR and Hall measurements were performed in temperatures from 4 K to 300 K. As discussed in Part B in the room temperature measurements, the polarity of MPE-induced AMR and SMR differs from each other only when the magnetic field is applied in *z* direction. This means that if the magnetic field is applied in *x* or *y* direction, even if both types are present at low temperature, it may still be difficult to separate them because of the same polarity. Therefore, we chose to apply the field in the out-of-plane direction (*z* direction) in the temperature dependent measurements. As our measurement system does not allow *in-situ* rotation of sample direction, for the temperature-dependent measurements in z-direction, we have changed the 5-nm-thick sample to samples with a thickness of 7 nm and 10 nm, respectively. As all the samples show similar MR and Hall behavior at room temperature, the change of samples will not compromise the consistency of discussion.

Fig. 3(c) is the out-of-plane MR of Ta (10 nm)/YIG at temperatures from 4 K to 300 K (similar results are obtained for Ta (7 nm)/YIG which are not shown here). The curves are clearly a superposition of two types of effects. For the curves at 10 - 300 K, a small positive MR is observed at large field region. This background MR comes from OMR as discussed in Part B, which is a result of the Lorentz force felt by the conduction electrons. Additionally, sharp central dips are observed in the range between -2000 Oe and 2000 Oe, which are consistent with the polarity and field range predicted by the SMR theory. This large positive MR is the result of the SMR caused by the change in YIG's magnetization direction below its saturation field. Therefore, the MR curves obtained at 10 - 300 K is dominantly a superposition of the OMR and SMR effect. For the curve at 4 K, the central dip polarity remains the same, indicating again the SMR contribution. However, the background MR polarity is flipped to be negative. This negative MR follows the polarity predicted by MPE, i.e., the AMR arises from the magnetized Ta layer, suggesting the large enhancement of MPE at low temperature. In this sense, the MR curve at 4 K is dominantly a superposition of the MPE-induced AMR and SMR effect. A schematic illustration of the above scenario is shown in Fig. 3(e). Same MR measurement was performed on Pt (5 nm)/Ta (5 nm) multilayer structure on $SiO_2$/Si to exclude the contribution from magnetic impurities in the as-sputtered Ta. The polarity remains the same for this sample at 4 K, indicating the flip in Ta/YIG is due to the presence of the ferrimagnetic YIG and MPE from YIG. Based on these different origins, the background MR is subtracted, and temperature-dependent SMR ratio is shown in Fig. 3(f) for both the 7 nm and 10 nm Ta samples. The size of the SMR ratio (~$10^{-5}$), is comparable to room temperature value for the 5 nm Ta sample. Despite some fluctuations, the overall trend for the SMR ratio is that it increases when the temperature and Ta thickness decrease. The former might be due to the enhancement of spin diffusion length at low temperature, while the latter confirms the nature of SMR as an interface effect.

Fig. 3(d) is the out-of-plane Hall resistance of Ta (10 nm)/YIG at temperatures from 4 K to 300 K (similar results are obtained for Ta (7 nm)/YIG which are not shown here). The shapes of the curves are similar to that of the 5 nm Ta sample at room temperature. After subtracting the linear ordinary Hall effect contribution, the Hall resistance ratio is shown in Fig. 3(g). The size (~$10^{-6}$) is one order of magnitude smaller than the SMR ratio, and has a similar temperature and thickness dependence as that of the SMR ratio, indicating the same origin for both MR and Hall resistance.

## IV. SUMMARY

In conclusion, unconventional MR effect was observed in Ta/YIG bilayer with a comparable size to Pt/YIG. Our MR and Hall results show that the electron transport property in the Ta overlayer is greatly influenced by YIG's magnetization, which could be explained by the SMR theory. Additionally, the observation of a flip of polarity in the temperature dependent MR measurements suggests that MPE-induced AMR is enhanced at low temperature. Our findings suggest that both MPE and SMR may contribute to the unconventional MR effect, in particular at low temperatures. A theory combining the features of MPE and SMR may be helpful in fully understanding the experimental results observed in NM/YIG bilayers.

## ACKNOWLEDGMENTS

This work was supported by the National Research Foundation of Singapore (Grants No. NRF-G-CRP 2007-05 and R-263-000-501-281). The authors wish to acknowledge Wei Zhang from National University of Singapore for her assistance during sample preparation, and Wei Ji, Meysam Sharifzadeh Mirshekarloo from Institute of Materials Research and Engineering (IMRE) for their assistances during structural characterization. K. Y. and S. S. wish to acknowledge the support from IMRE under project number IMRE/10-1C0107.

## FIGURE CAPTIONS:

FIG. 1. (a) Grazing incidence angle XRD pattern for the unpatterned reference Ta (10 nm)/YIG/GGG sample; (b) in-plane and out-of-plane M-H loop for YIG/GGG substrate. Inset in (a): XRD pattern for YIG/GGG wafer with (111) orientation; inset in (b): Superimposition of in-plane M-H loop and M-H loop obtained with a $90^{o}$ in-plane rotation of the sample with respect to its original position.

FIG. 2. (a)- (b) Schematic configuration for MR and Hall measurements; MR and Hall resistance measured with field applied in different directions for Ta (5 nm)/YIG: MR in x-direction (c), Hall effect in x-direction (d), MR in y-direction (e), Hall effect in y-direction (f), MR in z-direction (g), Hall effect in z-direction (h). An offset resistance of 3-4 mΩ is subtracted in Hall resistance. All measurements were performed at room temperature.

FIG. 3. (a)- (b) Schematic configuration for temperature-dependent MR and Hall measurements; (c) Temperature-dependence of MR with field applied in z direction for Ta (10 nm)/YIG; (d) Temperature-dependence of Hall resistance with field applied in z direction for Ta (10 nm)/YIG; (e) Schematic illustration for the change of MR origin at 4 K; (f) Temperature-dependence of SMR ratio for Ta (7 nm)/YIG and Ta (10 nm)/YIG; (g) Temperature-dependence of Hall resistance ratio for Ta (7 nm)/YIG and Ta (10 nm)/YIG. An offset resistance of 3-4 mΩ is subtracted in Hall resistance and the MR and Hall curves are vertically shifted for clarity.

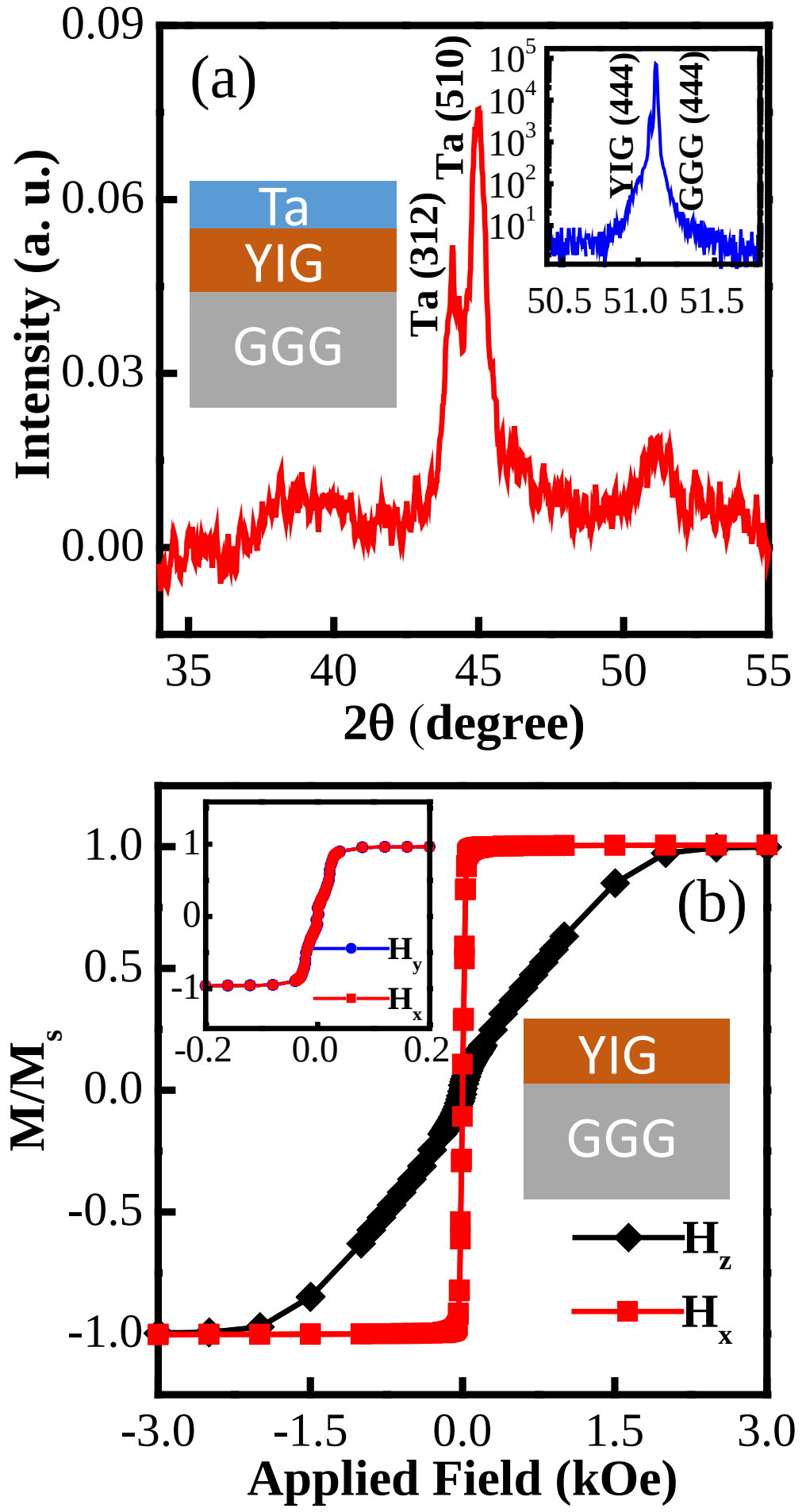

FIG. 1

Journal of Applied Physics

Yumeng Yang

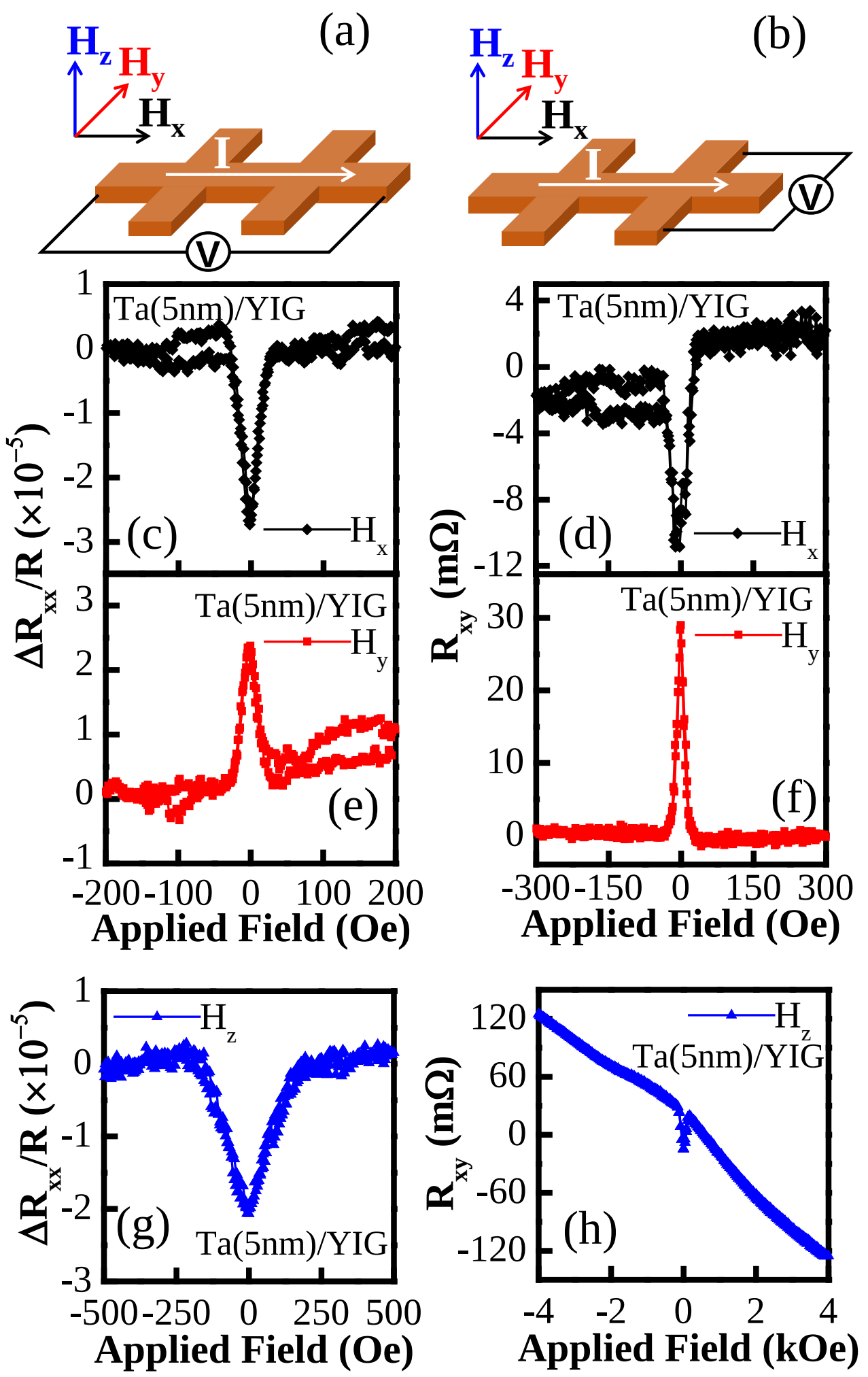

FIG. 2

Journal of Applied Physics

Yumeng Yang

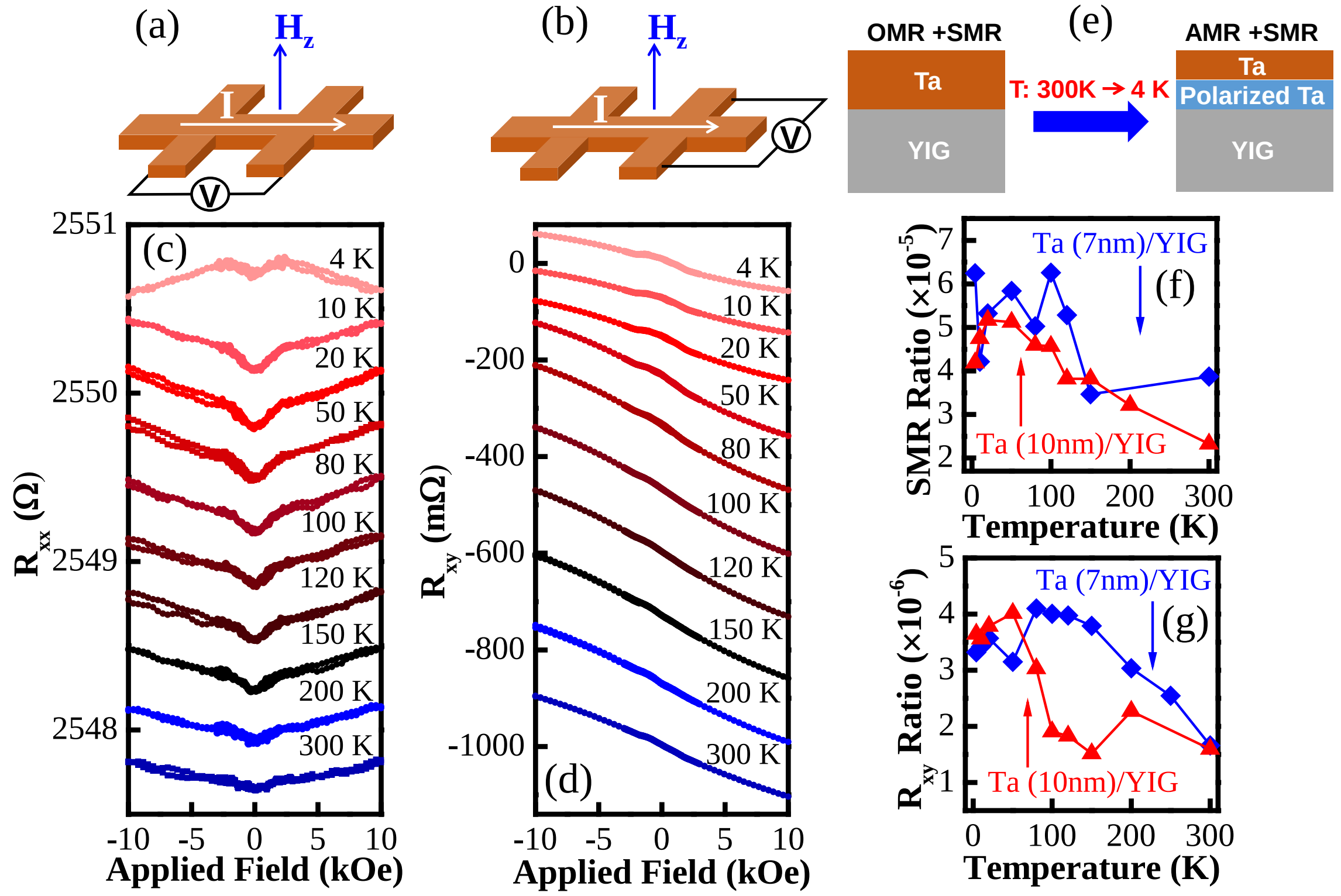

FIG. 3

Journal of Applied Physics

Yumeng Yang